# High-Tc superconductivity mechanism controlled by electric dipole correlation and charge correlation

**T. GUERFI***

*Department of physics. Faculty of Sciences.*
*M'hamed Bougara University, Boumerdes.*
*35000. Algeria*

---

**Abstract**

The model is based on parity and time broken symmetries phase transition leading to a particle antiparticle condensation. This approach gives a unified description of low and high $T_c$ superconductivity with a point of view differing from that of BCS theory. It is argued that high-temperature superconductors transform to an anti- ferroelectric state (long range order) prior to the onset of superconductivity, whereas in conventional materials, the material's crystal structure symmetry is the key to understand the mechanism of pairing by introducing a mirror plane polarization effect in the lattice. This effect is responsible for strong coupling and of two dimensionality aspect of HTSC, as it is described below.

---

*Author:

Name: GUERFI Tarek

Address: Laboratoire de Couches Minces et Interfaces, LCMI. Mentouri University. Constantine. 25000. Algeria

Tel/Fax: (00)213-31-81-86-64

*E-mail address:* tarek.guerfi @gmail.com

## 1. Introduction

The discovery of superconductivity at high temperature in hole-doped oxides in 1986 by Bednorz and Müller [1] caused a great debate among the scientific community. One of the first debate's questions asked was: are electron-phonons (e-p) interaction responsible for HTSC, as it has been proved for metallic superconductors? [2]. It is well known that the properties of high-temperature superconductors HTSC vary in an unusual way when a moderate density of holes is introduced into the material by chemical doping. Correlations show the parent undoped compound to be Mott insulator and, upon doping the underdoped compounds displays unusual metallic behaviour with increasing Tc. Doping beyond the optimal level yields normal metals with Fermi liquid behaviour and with decreasing Tc. This is one of the reasons why, despite large experimental and theoretical efforts, the nature of superconductivity in these materials remains unexplained. Many theories ascribing HTSC to a certain specific interactions, especially magnons, excitons, plasmons and bi-polarons or other interactions specified by some kind of symmetry [2, 3] have been suggested.

However, a large number of experimental and theoretical investigations indicate that two dimensionality aspects of the normal and superconducting state is one of the key factors in high temperature superconductivity [4, 5]. In Copper oxides (cuprates), it was found that the Ginzburg-Landau coherence length is less than the lattice spacing in the direction perpendicular to the $CuO_2$ planes (c-direction). Moreover, small buckling of these planes (on a scale of 1%) significantly reduces $T_c$ [6].

Superconductivity is of course a quantum phenomenon. However, classical arguments can shed important light into the fundamental physics of superconductors. In the sense of Bohr's correspondence principle, one can argue that the macroscopic quantum manifestations of superconductivity should be also understandable from a classical point of view [7]. In this paper, a new approach explaining the phenomenon of superconductivity, straightforwardly based on Coulombic framework interaction without invoking phonons, and leading to a unified description of low and high Tc superconductivity, with a point of view differing from that of BCS theory, is presented. It appears that the phenomenon of superconductivity is parity and time-reversal broken symmetries phase transition leading to particle-antiparticle condensation.

## 2. Structure of layered oxides:

The structure of one of the parent compounds of the HTSC, YBCO, is shown in fig.1 [8]. The material has a perovskite structure consisting of layers of CuO/BaO/$CuO_2$/Y/$CuO_2$/BaO/CuO along the *c*-axis. CuO is a one dimensional chain and $CuO_2$ is a two dimensional buckled plane. As the charge carriers are holes, $YBa_2Cu_3O_{7-\delta}$ is a `p'-type. In a stoichiometric orthorhombic unit cell ( $\delta$=0), with space group Pmmm, ideally, all the oxygen sites in the basal plane along the *a*-axis are empty and all the oxygen sites along the *b*-axis are 100% filled. In a non-stoichiometric orthorhombic compound, ($\delta$ is not equal to zero), the orthorhombicity is brought about by the unequal distribution of oxygen atoms in the basal (001) plane. All crystal structure of High-*Tc* compounds share a high symmetrical structure along the c axis (layered structure). An enormous amount of data on HTSC crystal structures has revealed that 'charge reservoir blocks' and 'active blocks' are alternately stacked along an unique direction in superconducting copper oxides. Accordingly, YBCO structure consists of BaO-CuO-BaO charge reservoir blocks and $CuO_2$-Y-$CuO_2$ active block. As schematically presented in fig.3 the HTSC compounds can be generally considered as an intergrowth of a charge reservoir blocks and active blocks. This motif occurs in all the high temperature superconducting copper oxides. An example of the most representative high-*Tc* compounds is given in table 1.

## 3. Mechanism of pairing

One may consider cuprates and the mechanism of pairing suggested unifies low and high-$T_c$ superconductivity. There is no doubt that superconductivity is based on pair of charge of magnitude 2e (Cooper pairs in case of BCS theory) or pairs of holes depending whether the materials is p or n type superconductor, with zero net momentum because the usual ac Josephson effect frequency *2eV/h* is observed [9], the observed flux quantum is of the usual size *hc/2e* [10] and Andreev reflection along time-reversed trajectories is seen [11] as with conventional superconductors. In fact, all superconductors exhibit four main properties: zero resistance, Meissner effect, flux quantization, Josephson effect, and gaps in the elementary energy excitation spectra. All this make one to think that the essential microscopic mechanism behind superconductivity is the same in all these materials in spite of their quite different natures.

As schematically shown in figure 3.a, HTSC compounds can be considered as an intergrowth of an active blocks and charge reservoir blocks. One of the best studied compounds is $YBa_2Cu_3O_7$. Its structure consists of BaO-CuO-BaO charge reservoir blocks and $CuO_2$-Y-$CuO_2$ active blocks. Once the material is cooled a spontaneous anti-ferroelectric (long range order) phase transition occurs in the active block at the Curie transition temperature. Figure 3.a shows schematically the electric dipole moments distribution between the mirror plane and the adjacent layers in the active blocks in the anti-ferroelectric state in case of YBCO compound. It must be noted that the mirror symmetry is not broken in the case there is no free charge carriers. High-temperature superconductors transform to an anti- ferroelectric state prior to the onset of superconductivity.

Indeed, an anti-polar phase transition occurs in the active block layers (a mirror electric dipole moment appears in the unit cell) at the Curie transition temperature. There are several studies that show the lattice geometry changes during the superconductivity transition which can be considered as the crystallographic signature of anti-ferroelectric phase transition in HTSC.

For YBCO material which is the best studied material, Srinivasan *et al* [12] claims that there is 0.4 °A anomalous jump in the *c* parameter (11.5 °A) of the lattice at the transition temperature whereas the variation in *a* and *b* parameters are negligible. Horn *et al* [13] has not observed a jump in the lattice parameters however, but he reports an anomaly in the orthorhombic splitting, *b-a*, near the superconducting transition whereas there is little or no anomaly in the unit cell volume. Also, in their paper, Schaefer et al [14] showed structural anomalies of $YBa_2Cu_3O_{6.9}$ at the superconducting transition temperature. A single phase specimen of superconducting $YBa_2Cu_3O_{6.9}$ has been investigated by these authors, using high resolution neutron powder diffraction in the temperature range from 16 to 300 K. Based on their result we were able to compute the dipole moment value upon the half of the YBCO unit cell and this gives a value of $1.1029 \times 10^{-29}$ C.m along the c axis (the polar axis). This provides the experimental evidence, in YBCO material case, of an antipolar (anti-ferroelectric) phase transition at the Curie temperature 92 K, which is the same superconducting transition temperature at optimal doping.

All high temperature superconductor material can be considered as an intergrowth of an anti-ferroelectric block (active blocks), and charge reservoir blocks since structural anomalies near Tc are a common feature of HTSC. C.Dong et al [15] reported structural anomalies occurring close to the superconducting phase transition in a nearly single phased $Bi_2Sr_2Ca_2Cu_3O$ sample, also superconducting single crystals of $Tl_2Ba_2CaCu_2O_8$ and $YBa_2Cu_4O_8$ showed anomalous structural change in the vicinity of the critical temperature [16]. Structural anomaly just above Tc was also observed in $(Hg_{1-x}Tl_x)Ba_2Ca_2Cu_3O_{8-\delta}$ superconductor [17], Qinlun Xu *et al* [18] have reported structural changes in single phase samples of Bi(Pb)-Sr-Ca-Cu-O at 110 K, and very recently similar to that in cuprates, lattice structure anomaly was reported in $LaFeAsO_{1-x}F_x$ at the onset superconducting transition temperature [19].

Dipole-dipole forces are short range and essentially anisotropic and their interaction can be either repulsive (positive dipole-dipole interaction energy) or attractive (negative dipole-dipole interaction energy). When a free charge carrier is introduced, which can be hole or electron depending on chemical doping, the parity symmetry is broken and a net Electric Dipole Moment (EDM) appear in the unit cell as consequence. It is noted that for single particle or a composite system the intrinsic EDM in non-degenerate state is also a signature of time-reversal broken symmetry.

This phase transition leads to the formation of perfectly correlated mirror charge carriers that are paired with the free charge carrier. The basic physics behind the mirror charge assumption is that the introduction of negative (or a positive charge) at the left or the right of the mirror plane breaks the mirror symmetry (electrical depoling effect) and lead to a formation of mirror charge on the other side of the mirror plane. In fact, a missed electron in the other side of the mirror plane which may restore the mirror symmetry is behind the formation of positive mirror charge. Thus an introduction of free electron in one side of the mirror plane is accompanied simultaneously with an introduction of a mirror charge in the other side of the mirror plane in the superconducting state. This situation is shown schematically in Figure 3.b. So, we have a case of a perfect screening of the free charge carrier and a formation of new state of matter. Even though, the Bose-Einstein condensate is neutral; the assumption of mirror charge does not violate the charge conservation principle. That is the divergence of the polarization (non uniform polarization) when the parity (mirror) symmetry is broken which supplies the electric charge of +e to the mirror charge, the material remains neutral.

$$\rho_b = -\nabla.\vec{P} \qquad (1)$$

where ***P*** is the net dipole polarization moment in the unit cell resulting from mirror symmetry breaking and $\rho_b$ is the bound charge (the mirror charge in this case) and we emphasize here that this is a perfectly real charge density.

It is well established that the underdoped HTC are characterized by the presence of a pseudogap in the normal-state excitation spectrum [20]. This pseudogap has been observed by a variety of transport, thermodynamic and spectroscopic probes [21] and is found to affect many normal-state properties.

So, one can speaks of two critical temperatures, a critical temperature $T_{c1}$ (Curie temperature) for the antipolar phase transition of the material and another critical temperature $T_{c2}$ for parity and time broken symmetries phase transition (particle- antiparticle condensation) controlled by free charge carriers density, both critical temperatures are the same at optimal doping.

However a depolarization may occurs above the Curie temperature $T_{c1}$ by thermal depoling, or by electrical depoling controlled by free charge density, or by mechanical depoling if the stress applied is too high. Thus an increase (decrease) in $T_{c1}$ and therefore to $T_{c2}$ may be possible by applying a pressure or increasing the free charge carriers density. Therefore, the bell-shaped curve relating the density of charge carriers to the superconducting critical temperature, seen experimentally, is explained.

## 4. The physical properties of free charge carrier images:

Superconductivity is a parity and time reversal broken symmetries phase transition leading to a particle-antiparticle condensation. The physical properties of antiparticles (mirror charge carriers) can be derived from CPT theorem which is a valid symmetry of nature, where C is charge conjugation, P is parity and T is time reversal.

A consequence of CPT invariance is that antiparticles can be treated mathematically as they were particles of the same mass and of oppositely signed charge of the same absolute value going backward in space and time. The CPT invariance means the following simultaneous operations:

- Charge conjugation (i.e. changing particle into antiparticle), $C\psi(r,t) = \bar{\psi}(r,t)$.
- Parity change (i.e. mirror reflection), $P\psi(r,t) = \psi(-r,t)$.
- Time reversal, $T\psi(r,t) = \psi(r,-t)$

The CPT invariance states that this does not change the physics, the wave function or in the language of field theory the field function $\psi(r,t)$ ,

$CPT\psi(r,t) = \bar{\psi}(-r,-t) \sim \psi(r,t)$.

$CPT\psi(r,t)$ represent a positron which can be considered in the Stückelberg-Feynman picture as a negative energy electron running backward in space-time.

## 5. The Hamiltonian operator of the particle–antiparticle pair:

When the parity symmetry is broken in the superconducting state, a net polarization develops in the unit cell. So, the simplest form of Landau theory for bulk ferroelectric starting from an expansion of the free energy in terms of the polarization, may also be applicable in case of superconductor material, accordingly:

$$F = \frac{1}{2}AP^2 + \frac{1}{4}BP^4 + \frac{1}{6}GP^6 \qquad (2)$$

where $A = A_0(T - T_{c2})$ (3)

and $T_{c2}$ is the critical temperature for mirror symmetry breaking phase transition. For second phase transition B > 0 and term in $P^6$ can be ignored.

In the other picture, we have a particle-antiparticle condensation. These particles, the particle (the free charge carrier) and its corresponding antiparticle, are characterized by CPT transformation. By applying C, P and T a single particle state is transformed into an antiparticle state with reversed spin and the same momentum.

The Hamiltonian operator of these pairs below the critical temperature can be written in the following quick form:

$$\hat{H} = \sum_k \varepsilon_k^p c_{k\uparrow}^+ c_{k\uparrow} + \sum_k \varepsilon_k^a \beta_{k\downarrow}^+ \beta_{k\downarrow} + \sum_{k',k} V_{k'k} c_{k'\uparrow}^+ \beta_{k'\downarrow}^+ \beta_{k\uparrow} c_{k\downarrow} \qquad (4)$$

Where $\varepsilon_k = \varepsilon_k^p = \varepsilon_k^a$ is the single particle (antiparticle) energy measured relative to Fermi level $\varepsilon_F$, and $c_{k\uparrow}^+$ $(c_{k\uparrow})$ is creation (annihilation) operator of particle and $\beta_{k\downarrow}^+$ $(\beta_{k\downarrow})$ is creation (annihilation) operator of antiparticle both satisfying the Fermi anticommutation rules:

$$\{c_{kS}, c_{k'S'}^+\} = \delta_{kk'}\,\delta_{SS'} \quad , \quad \{c_{kS}, c_{k'S'}\} = \{c_{kS}^+, c_{k'S'}^+\} = 0 \qquad (5)$$

$$\{\beta_{kS}, \beta_{k'S'}^+\} = \delta_{kk'}\,\delta_{SS'} \quad , \quad \{\beta_{kS}, \beta_{k'S'}\} = \{\beta_{kS}^+, \beta_{k'S'}^+\} = 0 \qquad (6)$$

The first (second) sum terms of the Eq (4) represents the total Kinetic energy of particles and their corresponding antiparticles. $V_{k'k}$ denotes the two body (particle–antiparticle) short range potential interaction arising from the Coulomb attraction between particle and its mirror image (antiparticle) . This interaction is attractive in the Columbic framework without invoking phonons.

Also, one notes that through the Stückelberg-Feynman picture (considering a positron as a negative energy electron running backward in space-time) one get essentially the same Hamiltonian as in Eq (24) of the original BCS paper [22].

## 6. Conventional superconductivity:

One can understand the mechanism of pairing in conventional superconductivity in terms of formation of a mirror charge carriers also (electron positron condensation). In conventional superconductivity, simple elements like Al, Pb, In, Sn, Tl, have the following crystal structure respectively, FCC/FCC/TET/TET/HEX and all possess the mirror symmetry. It is also, for compounds such as $Nb_3Sn$ $Nb_3Al$ or $V_3Ga$.

When the material is cooled, due to thermal contraction (expansion when heated), a weak coupling between planes, and especially the planes adjacent to the mirror planes, takes place. This effect occurs when, three atoms or ions taken respectively from: adjacent plane/ mirror plane/ adjacent plane, are brought together, the repulsion of ions charge distributions at sufficiently close separation, due to Pauli exclusion principle (short range interaction), induces a mirror electronic polarization of the ions adjacent to the mirror plane. An opposed electric dipole moments aligned perpendicular to the mirror plane appear between the mirror plane and the left and right adjacent planes. The strength of these dipoles is function of the electronic polarizability of the adjacent planes ions and of the interlayer distance which depend itself on thermal expansion coefficient of the material and of temperature. But this mechanism occurs only at low temperature, because the coupling effect between planes adjacent to the mirror plane cannot occur at high temperature. The thermal expansion coefficients for these materials are high, as a result the spacing between planes increases and the coupling decreases when the material is heated starting from the zero absolute temperature.

One can understand the difference in the critical temperatures for superconducting transition between metals, their alloys and ceramic oxides comes from the difference in the origin of the mirror polarisation. While it is ionic in case of HTSC (anti-ferroelectric phase transition which occurs relatively at high temperature), it is electronic in case of conventional superconductors.

Simple materials were reported to superconduct at surprisingly high critical temperatures 40 K for $MgB_2$ a binary compound [23] and 20K for lithium under pressure [24]. These results give more evidence that the mechanism behind superconductivity is the same in all types of superconductors.

## 7. Conclusion

Through a parity and time-reversal broken symmetry phase transition we get a unified description for low and high Tc superconductivity phenomenon. It is shown that high-temperature superconductors transform to an anti-ferroelectric state (long range order) prior to the onset of superconductivity. Whereas in conventional materials, the material's crystal structure symmetry is the key to understand the mechanism of pairing by introducing a mirror plane polarization effect in the lattice. In this model a particle-antiparticle pairs condensation, is responsible for the phenomenon of superconductivity in all type of materials.

## Acknowledgements

The author would like to thank his colleague M. IMADALOU for numerous discussions.

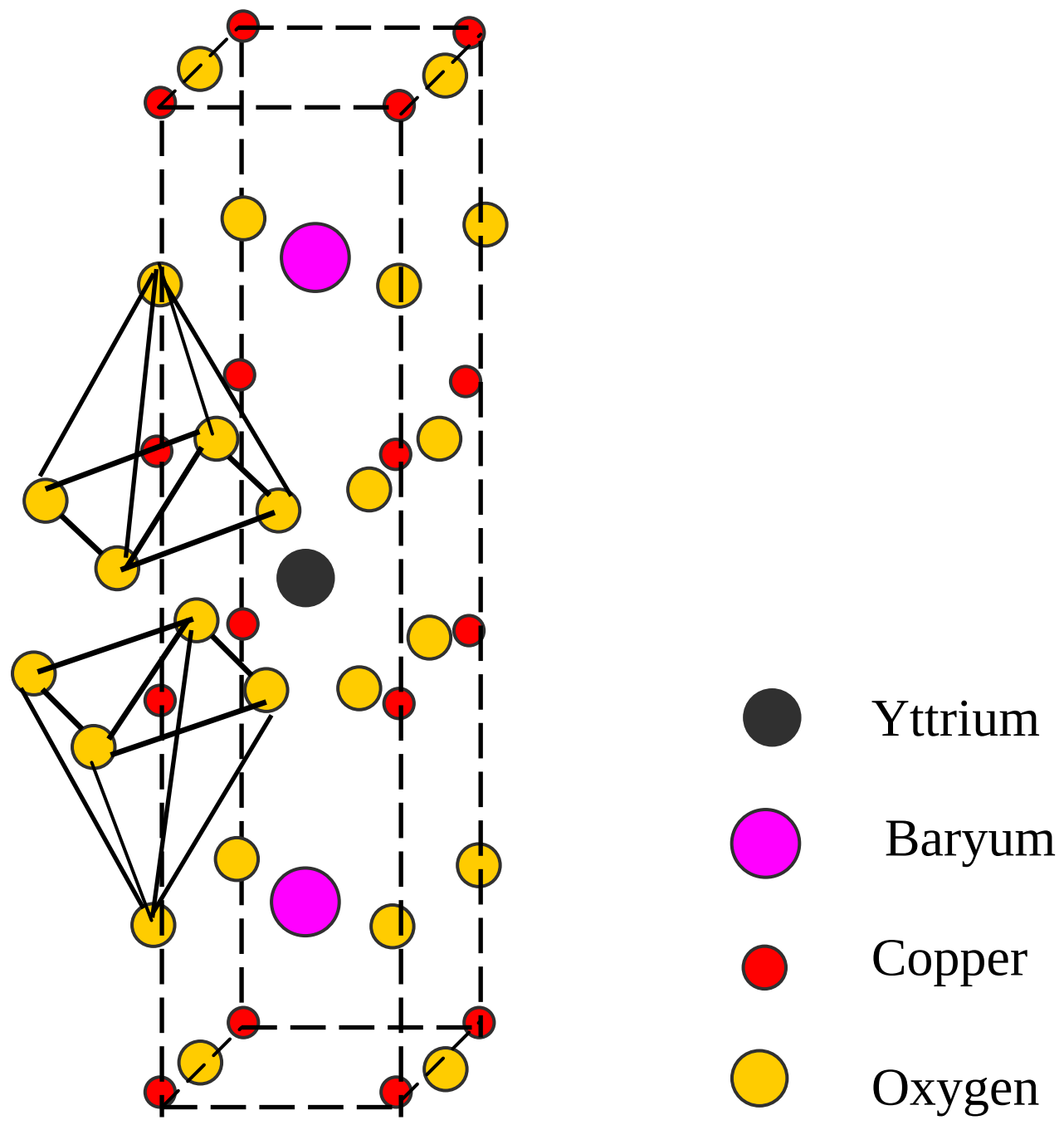
Yttrium
Baryum
Copper
Oxygen

Fig.1

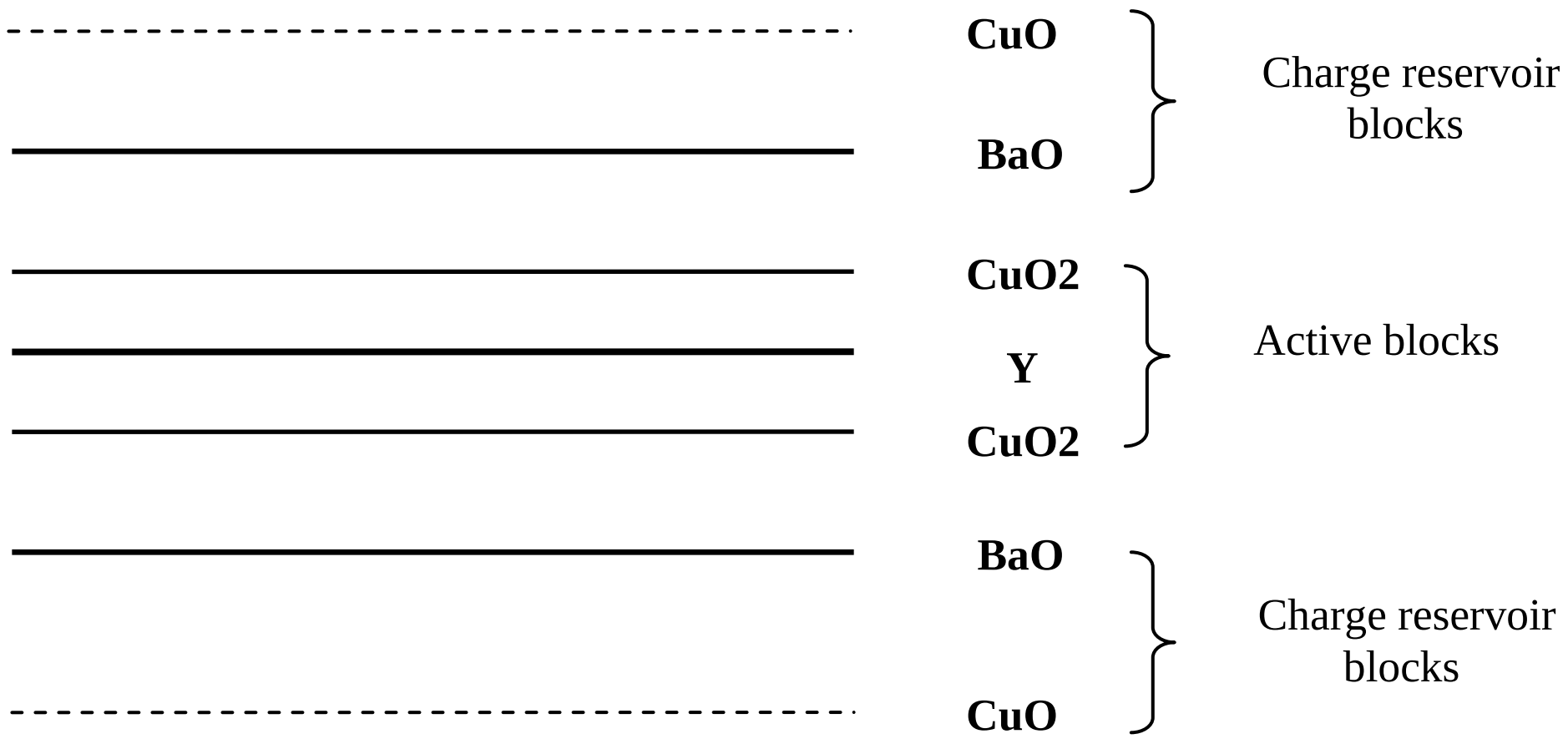

Fig.2

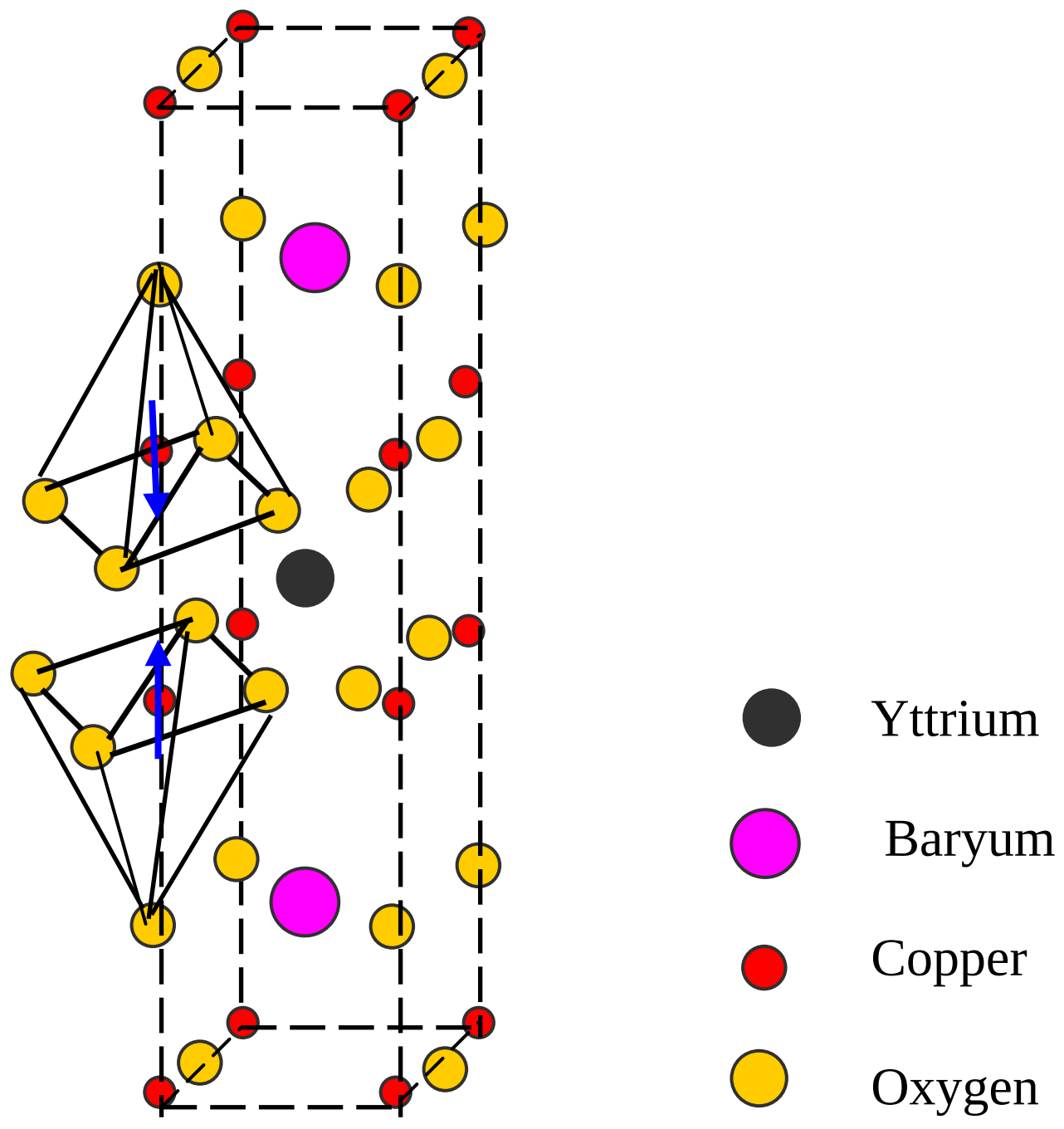
Yttrium
Baryum
Copper
Oxygen

Fig.3. a)

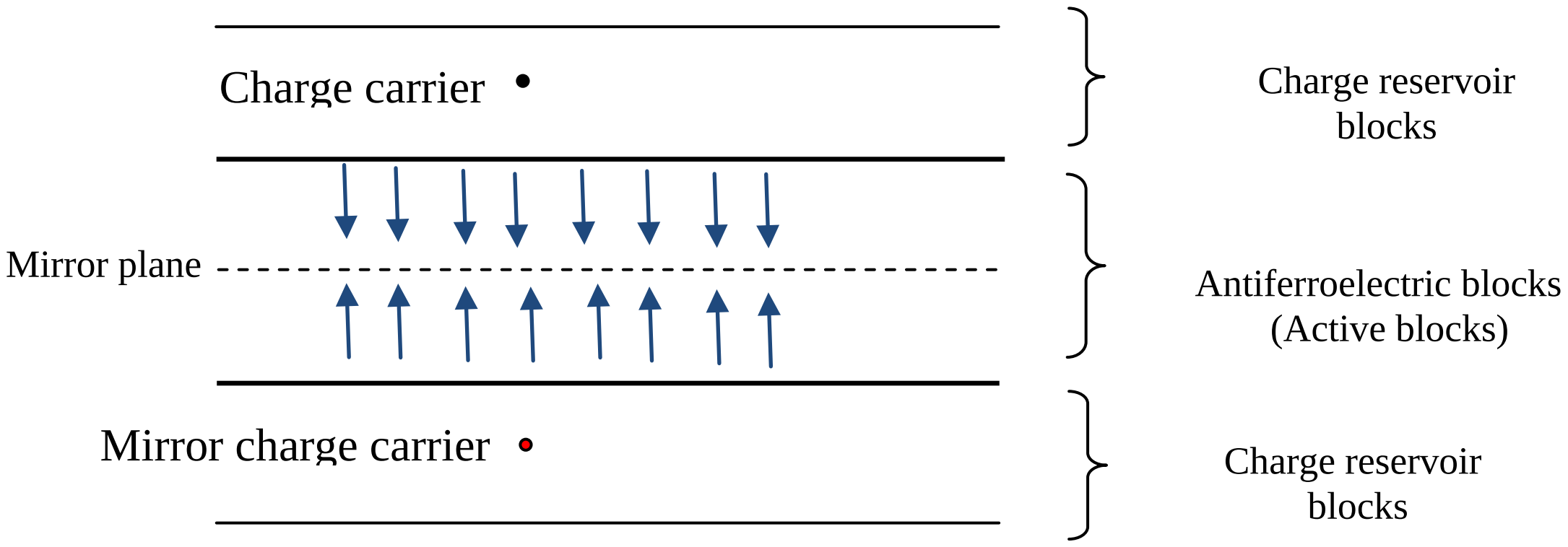

Fig.3. b)

| Compound | Tc(K) | Space group |
|---|---|---|
| $La_{2-x}M_xCuO_{4-y}$ | 38 | I4/mmm |
| M= Ba, Sr, Ca ; x~ 0.15, y is small | | |
| $Nd_{2-x}Ce_xCuO_{4-y}$( electron doped) | 40 | I4/mmm |
| $Ba_{1-x}K_xBiO_3$ ( isotropic, cubic) | 30 | I4/mmm |
| $Sr_{0.9}La_{0.1}CuO_2$ ( electron-doped infinite phase) | 43 | I4/mmm |
| $R_1Ba_2Cu_3O_7$ | 92 | Pmmm |
| R=Y, La, Nd, Sm, Eu, Er, Tm, Lu | | |
| $Bi_2Sr_2Ca_{n-1}Ca_nO_{2n+4}$ | | |
| n= 1('2201') | 10 | I4/mmm |
| n= 2 ('2212') | 85 | I4/mmm |
| n=3 ('2223') | 110 | I4/mmm |
| $Tl_2Ba_2Ca_{n-1}Cu_nO_{2n+4}$ | | |
| n=1('2201') | 90 | I4/mmm |
| n=2('2212') | 110 | I4/mmm |
| n=3('2223') | 125 | I4/mmm |
| $HgBa_2Ca_2Cu_3O_{10}$ | 133 | P4/mmm |

Table. 1. Some of the most representative high-Tc compounds.

**Figures captions:**

Fig.1 Crystal structure of YBCO compound.

Fig.2 Schematic picture of YBCO perovskite structure considered as a layered structure, or as an intergrowth of ‘charge reservoir blocks’ and ‘active blocks’.

FIG.3

a) Schematic picture of the anti-ferroelectric sate (long range order) occurring in the active blocks in YBCO unit cell. Note that the mirror symmetry (parity) is not broken in case there is no free charge carrier.

b) Schematic picture of parity symmetry breaking in the superconducting state, with presence of a free charge carrier and the formation of mirror charge carrier relative to the mirror plane.

**Table caption:**

Table .1 Some of the most representative high-Tc compounds.